# Discovering and Understanding 2D Semiconductors with High Intrinsic Carrier Mobility at Room Temperature

Chenmu Zhang, Ruoyu Wang, Himani Mishra, Yuanyue Liu*

*Texas Materials Institute and Department of Mechanical Engineering,*

*The University of Texas at Austin, Austin, Texas 78712, USA*

*Yuanyue.liu@austin.utexas.edu*

**Abstract:**

Two-dimensional (2D) semiconductors have demonstrated great potential for next-generation electronics and optoelectronics. However, the current 2D semiconductors suffer from intrinsically low carrier mobility at room temperature, which significantly limits their applications. Here we discover a variety of new 2D semiconductors with mobility one order of magnitude higher than the current ones and even higher than bulk silicon. The discovery is made by developing effective descriptors for computational screening of the 2D materials database, followed by high-throughput accurate calculation of the mobility using state-of-the-art first principles method that includes quadrupole scattering. Their exceptional mobilities are explained by several basic physical features. Particularly, we find a new feature: carrier-lattice distance, which is easy to calculate and correlates well with the mobility. Our work opens up new materials for high performance device performance and/or exotic physics, and improves the understanding of the carrier transport mechanism.

**Main text:**

One of the grand challenges for electronic materials research is to find an alternative to silicon with a suitable band gap, high carrier mobility at room temperature, and ambient stability, when thinning down to atomic thickness (for efficient gate control). The current candidates all suffer from one or more problems. For example, although graphene has very high carrier mobility, it does not have band gap. Two-dimensional (2D) crystalline semiconductors, on the other hand, currently suffer from intrinsically low carrier mobility at room temperature, due to strong scattering by phonons. For example, $MoS_2$, one of the most common 2D semiconductors, has an intrinsic electron mobility [1,2] < 200 $cm^2\ V^{-1}\ s^{-1}$, much lower than electron mobility of bulk silicon (1400 $cm^2\ V^{-1}\ s^{-1}$). Tremendous efforts have been devoted to search for higher-mobility 2D semiconductors. The past few years have witnessed the rise of 2D black phosphorus [3], indium selenide [4], etc., each of which has attracted great interest. However, their mobilities at atomic thickness are still not satisfactory (see Note S10 in the Supplementary Materials (SM) [5] and references [2,6-35] therein). Recent works cast doubt on the feasibility of realizing high mobility (e.g. > Si, 1400 $cm^2\ V^{-1}\ s^{-1}$) in atomically-thin semiconductors due to the limitation by dimensional effect [27,29].

Here we discover 13 monolayer semiconductors with mobility > 1400 $cm^2\ V^{-1}\ s^{-1}$, for example: BSb (electron mobility: ~ 5000 $cm^2\ V^{-1}\ s^{-1}$; hole: ~ 7000 $cm^2\ V^{-1}\ s^{-1}$), $ZrI_2$ (hole: ~ 5000 $cm^2\ V^{-1}\ s^{-1}$), $Sn_2H_2$ (~ 3000 $cm^2\ V^{-1}\ s^{-1}$), and $Ga_2Ge_2Te_2$ (electron: ~ 2000 $cm^2\ V^{-1}\ s^{-1}$). The discovery is made by search in the Computational 2D Materials Database (C2DB) [36,37]. We first formulated a set of descriptors based on the effective mass, Fröhlich scattering, and acoustic deformation potential scattering, and used them to narrow down the candidates. Then we accurately calculate their mobilities using state-of-the-art first principles method that includes quadrupole scattering [35]. To understand the origins of their high mobilities, we analyze the effects of electronic structure, density of scatterings [29], and electron-phonon coupling (EPC) strength for different phonon modes. We find that the high mobility can arise from small effective mass, high sound velocity, high optical phonon frequency, small ratio of Born charge vs. polarizability, and/or large "carrier-lattice distance" (a new physical feature to assess the EPC strength). By machine learning the relation between these features and the mobilities, we build a decision tree model to predict if a 2D semiconductor can have high mobility or not. The discovered materials as well as the mechanistic insights bring a step closer to the next-generation electronics/optoelectronics.

As shown in Fig. 1a, we first extract from the C2DB the materials with PBE band gap of 0.2-2 eV, resulting in 896 candidates. Then we select those labeled with "high dynamical stability", which gives 541 materials. Although their intrinsic mobilities can be accurately calculated using the Boltzmann transport theory in conjunction with density functional perturbation theory (DFPT) [35,38-41], it is computationally too expensive to calculate so many materials, many of which may have low mobility. To make the discovery more efficient, we first use several "descriptors" that can be easily calculated, to narrow down the candidates.

The first descriptor is the "combined effective mass" $M$, which we define as:

$$M = \sqrt{m_{\mathrm{t}}^* m_{\mathrm{d}}^*} \quad (1)$$

where $m_{\mathrm{t}}^*$ is the effective mass along carrier transport direction, and $m_{\mathrm{d}}^*$ is the density of state effective mass that can be approximated by $N\sqrt{m_x^* m_y^*}$ (here $N$ is the degeneracy of conduction/valence band extremes, and $x$ and $y$ are the two directions defined by the database). We use the combination of $m_{\mathrm{d}}^*$ and $m_{\mathrm{t}}^*$ because: (1) A lower $m_{\mathrm{d}}^*$ indicates a lower density of electronic states and thus less states available for carriers to be scattered to, which can increase the mobility as exemplified by $Sb_2$ [42] and $WS_2$ [43]. (2) When the scattering is fixed, decreasing $m_{\mathrm{t}}^*$ can further improve the mobility according to the Drude model; In order to evaluate the maximum mobility for anisotropic material, we use the smallest $m_{\mathrm{t}}^*$ available in the C2DB for a given material. Considering that $MoS_2$ electrons have a $M$ of 0.85, we use $M < 1$ as a criterion to screen the candidates with electron/hole mobility potentially higher than $MoS_2$ electron mobility. We find 149 (173) materials with electron (hole) $M$ satisfying this criterion.

The $M$ is a qualitative descriptor for mobility. To quantitatively estimate the mobility, we consider two important scatterings: Fröhlich scattering and acoustic deformation potential (ADP) scattering. The Fröhlich scattering originates from the interaction between carriers and the long-range dipolar potentials generated by optical phonons. It is the dominant scattering in many polar semiconductors [41,44]. We assume its $g$ has the form:

$$g_j^{\mathrm{F}}(\mathbf{q}) = i\frac{e^2}{2\Omega}\sum_{\kappa}\left(\frac{\hbar}{2M_{\kappa}\omega_j}\right)^{1/2}\frac{\mathbf{q}\cdot\mathbf{Z}_{\kappa}^{*}\cdot\mathbf{e}_{\kappa,j}(\mathbf{q})}{1+2\pi\alpha_{2\mathrm{D}}\,|\mathbf{q}|}, \tag{2}$$

where $\mathbf{q}$ is the phonon wavevector and $\hat{\mathbf{q}}$ is its unit vector, $\Omega$ is the area of unit cell, $\kappa$ is the index of atom in the unit cell, $M_\kappa$ is the atomic mass; $j$ is the index of the optical phonon mode, and $\omega_j$ is its phonon frequency (approximated by a constant obtained using the approach of Ref. [45]), $\mathbf{e}$ is the phonon eigenvector; $\mathbf{Z}^*$ is the Born effective charge, and $\alpha_{2\mathrm{D}}$ is the in-plane polarizability of the 2D crystal (approximated by $\alpha_{2\mathrm{D}} = (\alpha_{2\mathrm{D},xx}+\alpha_{2\mathrm{D},yy})/2$). See Note S3 in SM for calculation details. With the $g$ in hand, we can obtain the scattering rates via Eq. S3, and further the mobility due to the Fröhlich scattering ($\mu_{\mathrm{F}}$) via Eq. S1. This is done numerically, under the assumptions of a parabolic electronic band (with effective mass $m^* = (m_x^*+m_y^*)/2$) and multiple dispersion-less phonon modes (with frequency $\omega_j$).

The ADP scattering originates from the coupling of the electrons with LA phonons. The corresponding mobility can be estimated by [46]:

$$\mu_{\mathrm{ADP}} = \frac{e\hbar^3\rho v_{\mathrm{LA}}^2}{k_{\mathrm{B}}Tm_t^*\sqrt{m_x^*m_y^*}(D_{\mathrm{LA}})^2} = \frac{e\hbar^3 C_{2\mathrm{D}}}{k_{\mathrm{B}}Tm_t^*\sqrt{m_x^*m_y^*}(D_{\mathrm{LA}})^2}, \tag{3}$$

where $\rho$ is the area density, $v_{\mathrm{LA}}$ is the longitudinal sound velocity, and $C_{2\mathrm{D}}$ is the 2D elastic modulus (approximated by $C_{2\mathrm{D}} = (C_{2\mathrm{D},xx} + C_{2\mathrm{D},yy})/2$).

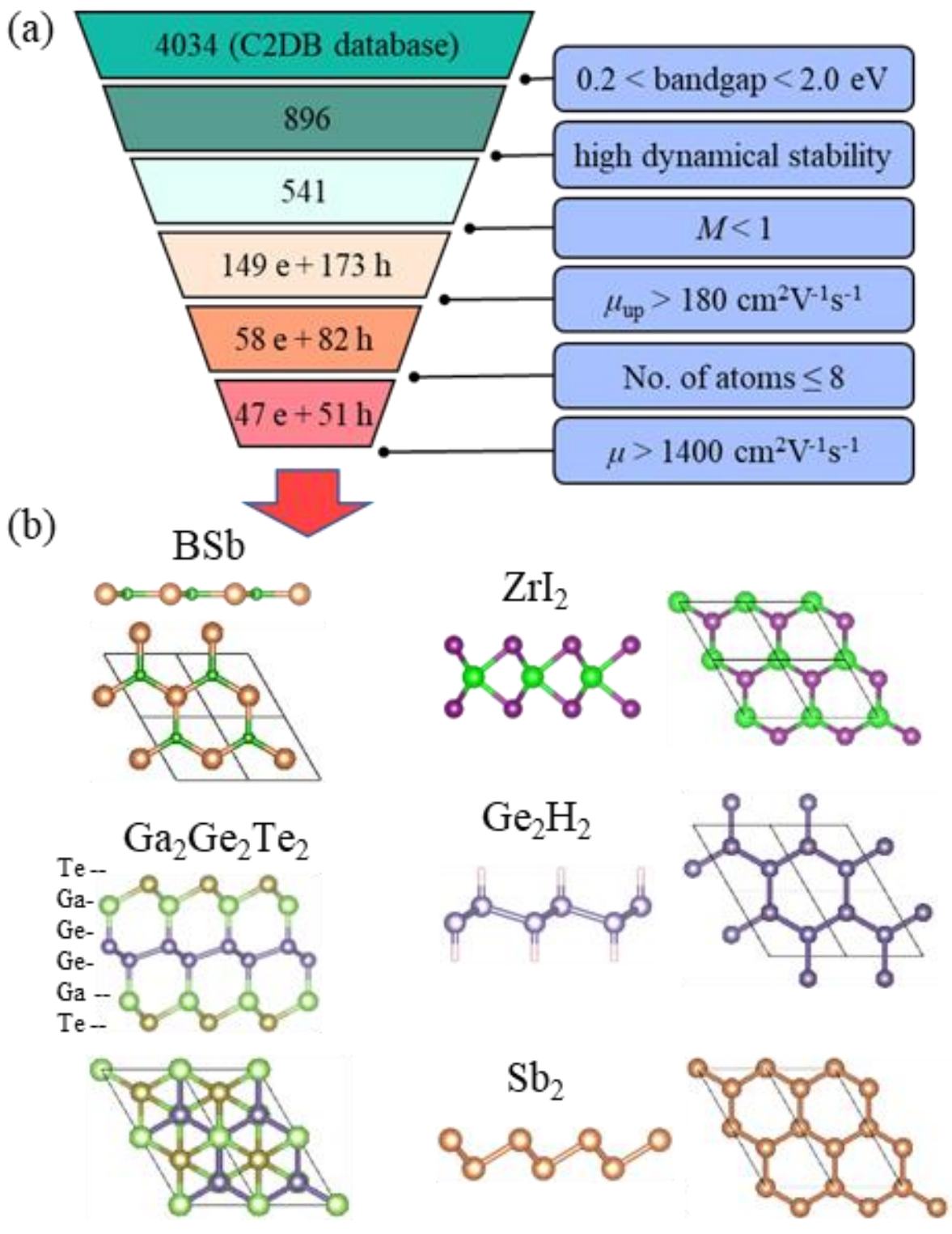

**Figure 1** (a) Illustration of screening procedures to discover 2D semiconductors with potential high carrier mobility. (b) Crystal structures for representative 2D semiconductors with mobilities over 1400

$cm^2\ V^{-1}\ s^{-1}$.

The mobility with both Fröhlich and ADP scatterings can be obtained from the individual mobilities following the Matthiessen's rule: $\mu_{up}^{-1} = \mu_F^{-1} + \mu_{ADP}^{-1}$. Since only two types of scatterings are considered, the $\mu_{up}$ should be regarded as the upper limit of the mobility, and can be used to exclude the low-mobility materials. Note that the physical parameters needed for $\mu_{ADP}$, $\mu_F$ and $\mu_{up}$ can be quickly extracted from the C2DB database, making them suitable for high-throughput screening. Although there are other scatterings (e.g. piezoelectric scattering, optical deformation potential scattering), their $g$ are relatively difficult to obtain, which makes it difficult to formulate the corresponding descriptors that can be easily calculated. Considering that the $\mu_{up}$ for $MoS_2$ electron is about 200 $cm^2\ V^{-1}\ s^{-1}$, we use $\mu_{up} > 180\ cm^2\ V^{-1}\ s^{-1}$ as a criterion and then obtain 58 (82) materials for electrons (holes).

After screening by $M$ and $\mu_{up}$ (see Fig. 1), we further exclude the materials with more than 8 atoms per unit cell which are computationally very expensive. Eventually, we accurately calculated 47 electron mobilities and 51 hole mobilities. Those results are shown in Fig. 2, together with the HSE band gaps. We also altered elements in the discovered high mobility materials, and found several additional high mobility semiconductors outside the database (see Note S6 in SM), which are also shown in Fig. 2. Particularly, we find the following 13 materials with mobility > 1400 $cm^2\ v^{-1}\ s^{-1}$, all having a hexagonal lattice (and thus isotropic mobilities) as shown in Fig. 1b: (1) III-V materials: BSb (e: 5167; h: 6935; 'e' for electron and 'h' for hole; in the unit of $cm^2\ V^{-1}\ s^{-1}$), AlBi (e: 2835; h: 3446), GaSb (e: 1809), BAs (e: 1524; h: 2439), InN (e: 2106) and BP (h: 1921). They all have an atomically flat structure like graphene. (2) $ZrI_2$ (h: 5138) and $HfI_2$ (h: 4782). Their structures are similar to that of 2H phase $MoS_2$ with the metal layer in the middle. (3) H functionalized IV materials: $Sn_2H_2$ (e: 3227; h: 2063) and $Ge_2H_2$ (e: 2791). (4) Group V materials: $Sb_2$ (h: 2044). (5) $Ga_2Ge_2Te_2$ (e: 1996) and $Al_2Ge_2Te_2$ (e: 2023). They have unique sextuple layered structure with atomic layers in order of Te-Ga(Al)-Ge-Ge-Ga(Al)-Te. Particularly, BSb, AlBi, $Sn_2H_2$ and BAs have both high electron mobility and hole mobility, as characterized by the ambipolar mobility $\mu_a$ (defined by $\mu_a = 2\mu_e\mu_h/(\mu_e+\mu_h)$ where $\mu_e$ and $\mu_h$ are electron and hole mobility respectively): 5922 for BSb, 3111 for AlBi, 2517 for $Sn_2H_2$ and 1876 for BAs. These excellent properties make them especially promising for electronic devices.

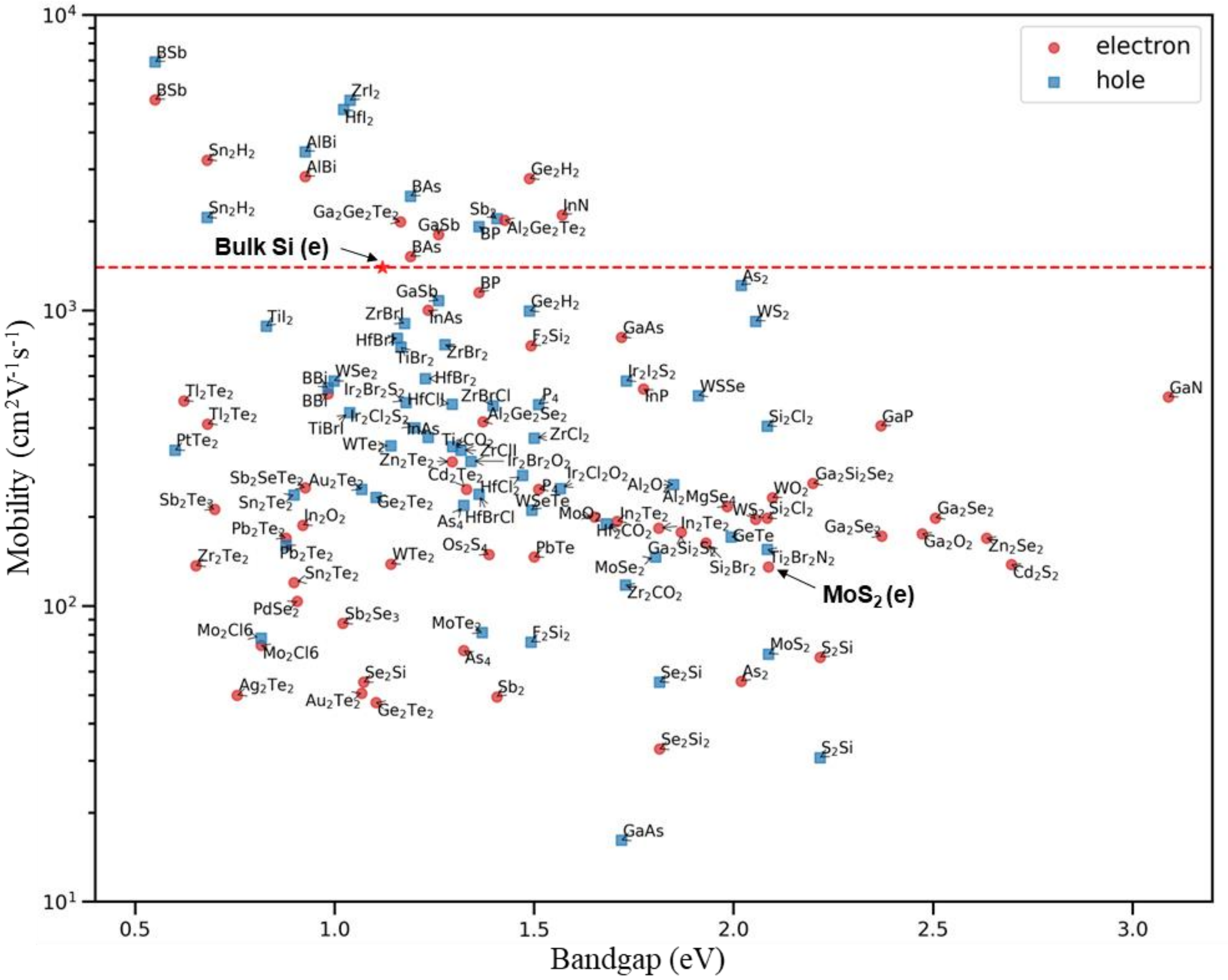

**Figure 2** Mobility vs. band gap (HSE) for various 2D semiconductors. For comparison, the electron mobilities for $MoS_2$ and bulk Si are marked.

In order to understand why the mobilities are so high for those materials, we calculate the “Drude effective mass” ($\bar{m}^*$) and “Drude scattering rate” ($1/\bar{\tau}$) (see Note S7 in SM for definitions) [42]. The $\bar{m}^*$ is fully determined by the electronic structure and its occupation, while the information about phonons and the EPCs are wrapped in $1/\bar{\tau}$. Fig. 3a shows the $1/\bar{\tau}$ vs. $\bar{m}^*$ for the materials with e/h mobility > 1400 $cm^2$ $V^{-1}$ $s^{-1}$, as well as those for $MoS_2$ (e) and Si (e) for comparison. We find that they all have $\bar{m}^*$ smaller than $MoS_2$, and $1/\bar{\tau}$ lower than $MoS_2$, which together give rise to their higher mobilities. However, the contributions from $1/\bar{\tau}$ and $\bar{m}^*$ are different. For example, the $1/\bar{\tau}$ of $Ga_2Ge_2Te_2$ electron is only 50% of $MoS_2$ (15 vs. 30 $ps^{-1}$), while its $\bar{m}^*$ is 14% of $MoS_2$ (0.058 vs. 0.43 $m_e$), which is the major contributor to its ~ 15 times higher mobility than $MoS_2$ (1996 vs. 136 $cm^2$ $V^{-1}$ $s^{-1}$). In contrast, $ZrI_2$ has a comparable $\bar{m}^*$ (0.40 $m_e$) to that of $MoS_2$ (0.43 $m_e$), but its $1/\bar{\tau}$ is much lower (0.85 vs. 30 $ps^{-1}$), resulting in 38 times higher mobility than $MoS_2$ (5138 vs. 136 $cm^2$ $V^{-1}$ $s^{-1}$). BSb has both a small $\bar{m}^*$ (0.09 $m_e$ for hole and electron) and a low $1/\bar{\tau}$ (2.8 $ps^{-1}$ for hole and 3.6 $ps^{-1}$ for electron), which together make it the highest mobility material. When comparing with bulk Si, we find that the main reason why $ZrI_2$ has a higher mobility is the lower $1/\bar{\tau}$, while for BSb and $Ga_2Ge_2Te_2$, it is the $\bar{m}^*$. Note that it is unusual to have low $1/\bar{\tau}$ for 2D semiconductors compared with Si, as the dimensionality effect results in a high "density of scatterings" (see Ref. [29] and Note S12 in SM). However, if the EPC is sufficiently weak, it is possible to achieve lower $1/\bar{\tau}$ and thus higher mobility than Si, as exampled by $ZrI_2$ (see Fig. S5 for details).

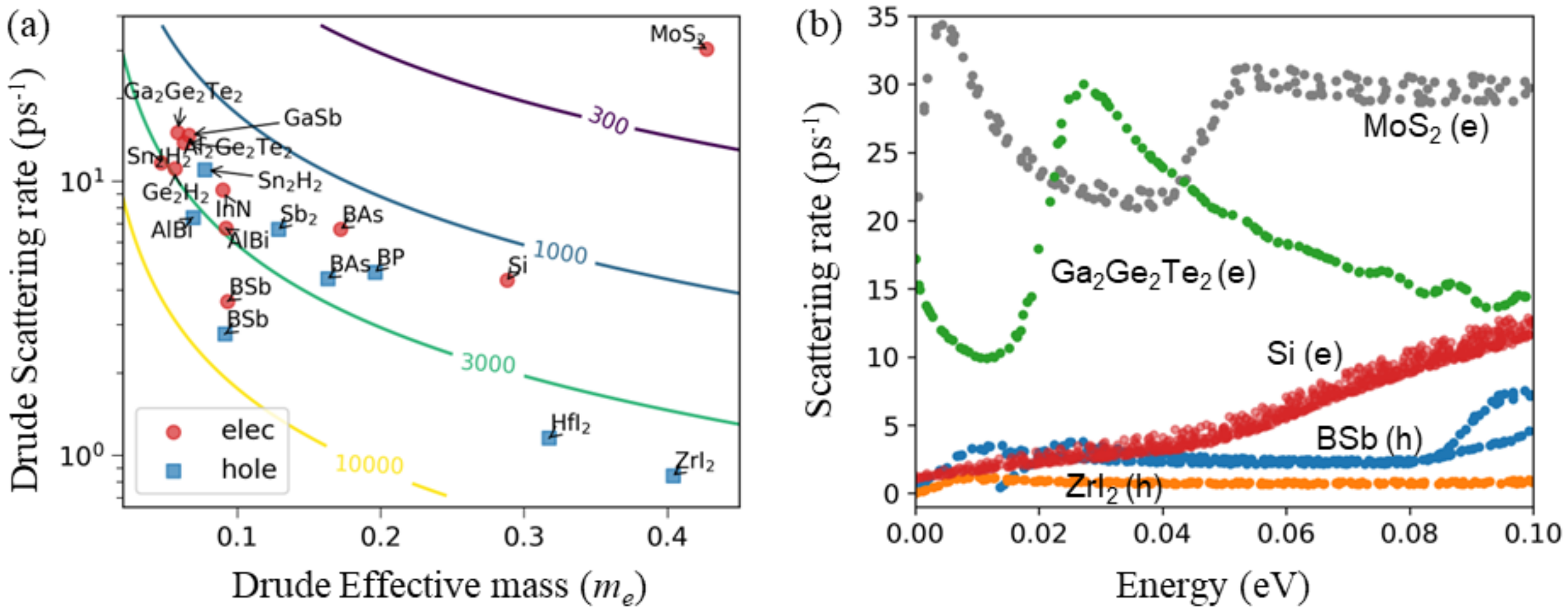

**Figure 3** (a) Drude scattering rate and Drude effective mass for high-mobility (> 1400 cm$^2$ V$^{-1}$ s$^{-1}$) 2D semiconductors. The lines are the iso-mobility contours. (b) Scattering rates of three representative high-mobility materials. For comparison, $MoS_2$ and Si data are also shown.

To further understand why scattering rate can be so low in some materials, we focus on 3 representative materials: BSb, $ZrI_2$, and $Ga_2Ge_2Te_2$. We choose them because: (1) BSb and $ZrI_2$ are the first two highest mobility 2D semiconductors; (2) There exists vdW layered bulk $Ga_2Ge_2Te_2$, suggesting that its 2D form may be easy to experimentally realize [47,48]. Fig. 3b compares the state-dependent scattering rates for these 3 materials and $MoS_2$, which shows the order: $MoS_2 > Ga_2Ge_2Te_2 \gg$ BSb > $ZrI_2$, consistent with that seen from $1/\bar{\tau}$. To gain further insights, we decompose the scattering rates into contributions from individual phonon modes. We focus on the LA and LO modes, as they are often the dominant scattering sources to the intrinsic mobility. Figure S3 compares the mode-resolved scattering rates for different materials. Interestingly, compared with $MoS_2$, $Ga_2Ge_2Te_2$ has a stronger LO scattering, while a weaker LA scattering. For BSb and $ZrI_2$, the scatterings are weaker for both LO and LA modes.

By analyzing the mode-resolved "density of scatterings" [29] and the EPC strength (see Note S8 in SM), we can explain the scattering rates using basic physical features. For example, the stronger LO scattering in $Ga_2Ge_2Te_2$ than that in $MoS_2$ is due to the larger ratio of Born charge to the in-plane polarizability ($R_{B/P}$), which gives a stronger EPC for the long-range dipolar perturbation potential induced by the LO phonons. In contrast, the Born charge vanishes in $ZrI_2$, resulting in a weak LO scattering. Although BSb has a large ($R_{B/P}$) and thus a strong LO EPC, the population of LO phonons is limited because of the high LO frequency ($\omega_{LO}$), therefore it also has a weak LO scattering. Interestingly, $Ga_2Ge_2Te_2$, BSb and $ZrI_2$ all have weak LA EPCs, which leads to their weak LA scatterings (additionally, BSb has a high sound velocity and hence few LA phonons, which also contributes to its weak LA scattering). In order to obtain an intuitive understanding of the LA EPC strength, we propose a new, simple yet effective feature: carrier-lattice distance $d_{c\text{-}l}$, defined as:

$$d_{c\text{-}l}(\text{CBM/VBM}) = \int_{uc} d\mathbf{r}\, |\psi_{\text{CBM/VBM}}(\mathbf{r})|^2 \min_\alpha\{|\mathbf{r} - \mathbf{R}_\alpha|\} \quad (4)$$

where the CBM/VBM indicates the electronic state at conduction band minimum or valence band maximum, the $\psi$ is the corresponding wavefunction, $\mathbf{R}_\alpha$ is the position of nucleus $\alpha$, and uc denotes the unit cell. This new feature $d_{c\text{-}l}$ quantifies the distance between the carrier (represented by the CBM/VBM) and the lattice, as illustrated in Fig. 4f. Since the perturbation induced by lattice displacement is generally weaker in the region farther from the nuclei, it is intuitive to expect that a larger $d_{c\text{-}l}$ will result in a weaker EPC.

To see more directly how the mobility is correlated by those basic physical features, we plot the mobility vs. $M$, $R_{B/P}$, $v_{LA}$, $\omega_{LO}$, and $d_{c\text{-}l}$ for all the materials in Fig. 4. It shows a small $M$, small $R_{B/P}$, high $v_{LA}$, high $\omega_{LO}$, and a large $d_{c\text{-}l}$ can benefit the mobility. By machine learning these correlations, we build a decision tree model (see Note S9 in SM) to quantify the common features shared by the high mobility (> 1400 cm$^2$ V$^{-1}$ s$^{-1}$) materials. We find that most of them have $M < 0.474$, $n_{LO}\, R^2_{B/P} < 0.066$ Å$^{-1}$ (where $n_{LO}$ is the number of LO phonons), $d_{c\text{-}l} > 1.11$ Å and $\omega_{LO} > 15$ meV. These criteria can help quickly assess the mobility of 2D semiconductors.

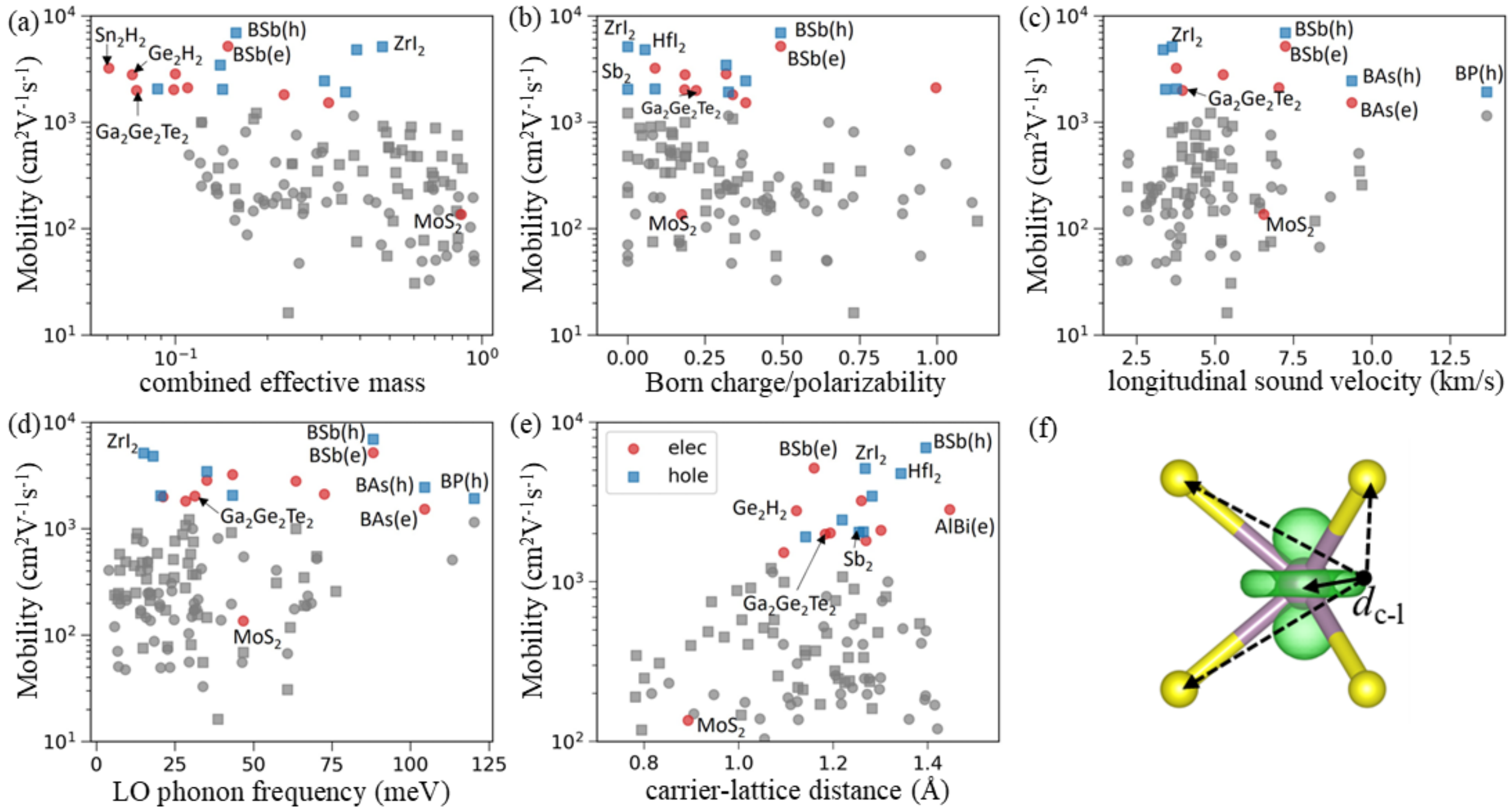

**Figure 4** Mobility vs. various basic physical features: combined effective mass (a), ratio of Born charge to in-plane polarizability (b), longitudinal sound velocity (c), and LO phonon frequency (d), and carrier-lattice distance (e; see text for definition and f for illustration). The high-mobility (> 1400 cm$^2$ V$^{-1}$ s$^{-1}$) 2D semiconductors are highlighted by red (for electron) and blue (for hole).

In summary, by developing effective descriptors for computational screening followed by high-throughput accurate calculation of the mobility, we discovered a number of high-mobility (even higher than bulk sillicon) 2D semiconductors, such as BSb, $ZrI_2$, $Sn_2H_2$ and $Ga_2Ge_2Te_2$. Their extraordinary mobilities are explained by basic physical features, including small effective mass, high sound velocity, high optical phonon frequency, small ratio of Born charge to polarizability, and/or large carrier-lattice distance. The feasibility of synthesizing those materials is discussed in Note S11 in the SM. We expect our predictions would stimulate experimental realizations, and the insights offered in this work may lead to further discoveries.

**Acknowledgements:**

This work is supported by Welch Foundation (F-1959-20210327) and NASA (80NSSC22K0264). The calculations used computational resources at Texas Advanced Computing Center (TACC), National Renewable Energy Laboratory (NREL), and Anvil through allocation CHE190065 from the Advanced Cyberinfrastructure Coordination Ecosystem: Services & Support (ACCESS) program.